  \documentclass{cjour}

\newfont{\Bb}{msbm8 scaled\magstep{1}}
\newcommand{\rc}{\mbox{\Bb R}}

\usepackage{graphics}
\usepackage{amsmath}
\numberwithin{equation}{section}
\begin{document}








\title
{Identification of Multilayered Particles from Scattering Data by a Clustering Method
}
\author{S. Gutman}

\affil{Department of Mathematics, University of Oklahoma, Norman, OK 73019, USA}

\email{sgutman@ou.edu}

\abstract{
A multilayered particle is illuminated by plane acoustic
or electromagnetic waves of one or several frequencies. We consider the
inverse scattering problem for the identification of the layers and of the
refraction coefficients of the scatterer in a non-Born region of scattering. 
Local deterministic and global probabilistic minimization methods
are studied. A special Reduction Procedure is introduced to reduce 
the dimensionality of the minimization space. Deep's  and the Multilevel
 Single-Linkage methods for global minimization are used for the solution of the inverse problem.
Their performance is analyzed for various multilayer configurations.
}



\begin{article}

\section{Introduction}

Many practical problems require an identification of the internal
structure of an object given some measurements on its surface. In this
paper we study such an identification for a multilayered particle
illuminated by acoustic or electromagnetic plane waves.
Thus the problem discussed here is an inverse scattering problem. A similar 
problem for the particle identification from the light scattering data is studied in \cite{zub}.
The
precise formulation of the problem is postponed till Section 2. Our approach is to reduce
the inverse problem to the best fit to data multidimensional
minimization. This is done in Section 3. It is also shown there that more than one
frequency of the incoming waves is required to provide a stable
identification. The resulting minimization is a challenging problem,
since
the objective function
has many narrow local minima. Finding a global
minimum (the sought identification) is the main subject of the study
here. In Section 4 we analyze various local minimization methods and
develop a special Local Minimization Method. This method, together with a
specially designed Reduction Procedure, is capable of finding 
this type of local minima. In Section 5 
  Rinnooy Kan and Timmer's 
Multilevel Single-Linkage Method for global minimization is presented.
It is paired with the Local
Minimization Method of Section 4, and, finally, gives the tool for the successful
scatterer's identification. A detailed numerical evidence of the
performance of this method is presented in Section 6.

\section{Direct Problem}
Let $D\subset \rc^2$ be the circle of a radius $R>0$, 
\begin{equation}\label{21}
D_m=\{x\in \rc^2\, :
r_{m-1}<|x|<r_m\,,\quad m=1,2,\dots ,N\}
\end{equation}
and $S_m=\{x\in \rc^2 : |x|=r_m\}$ for $0=r_0<r_1<\cdots<r_N < R$. 
Suppose that a multilayered scatterer in $D$ has a constant refractive
index $n_m$ in the region $D_m\,,\quad m=1,2,\dots ,N $. If the
scatterer is illuminated by a plane harmonic wave 
then, after the time dependency is eliminated, the total field
$u(x)=u_i(x)+u_s(x)$  satisfies the Helmholtz equation
\begin{equation}\label{22}
\Delta u+k_0^2u=0\,,\quad |x|>r_N
\end{equation}
where $u_i(x)=e^{ik_0x\cdot\alpha}$ is the incident field and $\alpha$ is
the unit vector in the direction of propagation. The scattered field
$u_s$ is required to satisfy the Sommerfeld radiation condition at infinity, 
see \cite{coltonkress}.

Let $k_m^2=k_0^2n_m$.
We consider the following transmission problem 

\begin{equation}\label{23}
\Delta u_m+k_m^2u_m=0\,\quad x\in D_m\,,
\end{equation}

under the assumption that the fields $u_m$ and their normal derivatives are
continuous across the boundaries $S_m\,, \,m=1,2,\dots ,N$.

In fact, the choice of the boundary conditions on the boundaries $S_m$
depends on the physical model under the consideration. The above model
may or may not be adequate for an electromagnetic or acoustic
scattering, since the model may require additional parameters (such as
the mass density and the compressibility) to be accounted for. However,
since the goal of this paper is to study algorithms capable to resolve
the Inverse Scattering Problem, we will accept the above simplified
problem here. For more details on transmission problems, including the 
questions on the existence and the uniqueness of the solutions,  see
\cite{sabatier}, \cite{athrammstrat} and \cite{ewing}.

The Inverse Problem to be solved is:

{\bf IPS:} {\it Given $u(x)$ for all $x\in S=\{x: |x|=R)$
at a fixed $k_0>0$, find
the number $N$ of the layers, the location of the layers, and their refractive indices
$n_m \,, \,m=1,2,\dots ,N$ in \eqref{23}.
}

Here IPS stands  for a Single frequency Inverse Problem.
 Numerical experience shows that there
are some practical difficulties in the successful resolution of the IPS
even when no noise is present.
While there are some results on the uniqueness for the IPS (see
\cite{athrammstrat}), assuming that the refractive indices are known, and only
the layers are to be identified, no stability estimates are available.
The identification is successful, however, if the scatterer is subjected 
to a probe with plane waves of several frequencies. Thus we state the Multifrequency
Inverse Problem:

{\bf IPM:} {\it Given $u^p(x)$ for all $x\in S=\{x: |x|=R)$
at a finite number $P$ of wave numbers $k_0^{(p)}>0$, find
the number $N$ of the layers, the location of the layers, and their refractive indices
$n_m \,, \,m=1,2,\dots ,N$ in \eqref{23}.
}

\section{Best Fit Profiles}
If the refractive indices  $n_m$ are sufficiently close to $1$, then we say that the
scattering is weak. In this case the scattering is adequately described by the Born 
approximation, and there are methods for the solution
of the above Inverse Problems. See \cite{coltonkress},\cite{coltonmonk},
\cite{r2} and \cite{r3} for further
details. However, when such an assumption is inappropriate, the
preferable method is to match the given observations to a set of
solutions for the Direct Problem. Since our interest is in the solution of
the IPS and IPM in the non-Born region of scattering, we choose to
follow the best fit to data approach. This approach is used widely in a
variety of applied problems, see e. g. \cite{biegler}.

Note, that, by the assumption, the scatterer has the rotational
symmetry. Thus we only need to know the data for one direction of the incident
plane wave. For this reason we fix $\alpha=0$ in (2.2) and assume that
the (complex) data functions 

\begin{equation}
g^{(p)}(\theta)\,,\quad p=1,2,\dots ,P
\end{equation}

are given for $0\leq \theta< 2\pi$, corresponding to the observations
measured on the surface $S$ of the ball $D$ for a finite set of free space
wave numbers $k_0^{(p)}$.

Fix a positive integer $M$. Given a configuration

\begin{equation}
Q=(r_1,r_2,\dots,r_M,n_1,n_2,\dots,n_M)
\end{equation}

we solve the Direct Problem (2.2)-(2.3) (for each free space wave number  $k_0^{(p)}$)
 with the layers 
$D_m=\{x\in \rc^2\, :
r_{m-1}<|x|<r_m\,,\quad m=1,2,\dots ,M\}$, and the corresponding refractive indices
$n_m$, where $r_0=0$.
Let
\begin{equation}
w^{(p)}(\theta)=u^{(p)}(x)\big|_{x\in S}\,.
\end{equation}

Fix a set of angles $\Theta=(\theta_1,\theta_2,\dots,\theta_L)$
and  let

\begin{equation}
\|w\|_2=(\sum_{l=1}^L w^2(\theta_l))^{1/2}
\end{equation}

Define
\begin{equation}
\Phi(r_1,r_2,\dots,r_M,n_1,n_2,\dots,n_M)=\frac{1}{P}\sum^P_{p=1}\frac{\|w^{(p)}-
g^{(p)}\|_2^2}{\|g^{(p)}\|_2^2}\,,
\end{equation}

where the same set $\Theta$ is used for $g^{(p)}$ as for $w^{(p)}$.

We solve the IPM by minimizing the above best fit to data functional $\Phi$
over an appropriate set of admissible parameters $A_{adm}\subset
\rc^{2M}$.

It is reasonable to assume that the underlying physical problem gives
some estimate for the bounds $n_{low}$ and $n_{high}$
of the refractive indices $n_m$ as well as for the bound $M$ of
the expected number of layers $N$. Thus, 

\begin{equation}
A_{adm} \subset \{(r_1,r_2,\dots,r_M,n_1,n_2,\dots,n_M)\ : \ 0\leq r_i\leq R\,,\ 
n_{low}\leq n_m \leq n_{high}\}\,.
\end{equation}

Note, that the admissible configurations must also satisfy

\begin{equation}
r_1\leq r_2\leq r_3 \leq\dots\leq r_M\,.
\end{equation}

As it was already mentioned in Section 2, the numerical evidence shows
that IPS is, practically, unresolvable. Here is an example to illustrate
the situation. Let the configuration $Q_1$ be 
 $(0.4,0.6,0.49,9.0)$ with $N=2$ and
$R=1.0$.
Thus $Q_1$ corresponds to the two layer cylinder

\[
n(x)=\begin{cases} 0.49 & 0 \leq x < 0.4\\
9.0 & 0.4\leq |x| < 0.6\\
1.0 & 0.6\leq |x| \leq 1.0
\end{cases}
\]

Let $Q_2=(0.3794,0.5662,0.6377,0.040,8.282,5.969)$ with $N=3$ and $R=1.0$,
thus $Q_2$ corresponds to the three layer cylinder

\[
n(x)=\begin{cases} 
0.040 & 0\leq |x| < 0.3794\\
8.282 & 0.3794\leq |x| < 0.5662\\
5.969 & 0.5662\leq |x| < 0.6377\\
1.0 & 0.6377\leq |x| \leq 1.0
\end{cases}
\]

Let the data $g(\theta)$ be collected for just one wave number
$k_0=3.0$. Figures 1 and 2 show the real and imaginary parts of the
solutions for these two configurations. The solutions are practically
indistinguishable, especially if noise is  present. Letting $Q_1$ to be the original 
configuration for
which the data $g(\theta)$ is observed, the value of $\Phi$ at the configuration $Q_2$
is just 0.00012. Thus, there is
no way (by any method) to determine the original configuration $Q_1$ of the scatterer.
 Clearly, there are many more configurations
that would produce practically identical observations. Even if it could be proven
that, theoretically, there is a unique solution for this IPS, it would
be useless in practice, because of this and other practically undistinguishable
configurations. 

\begin{figure}[tb]
\vspace{5pc}
\includegraphics*{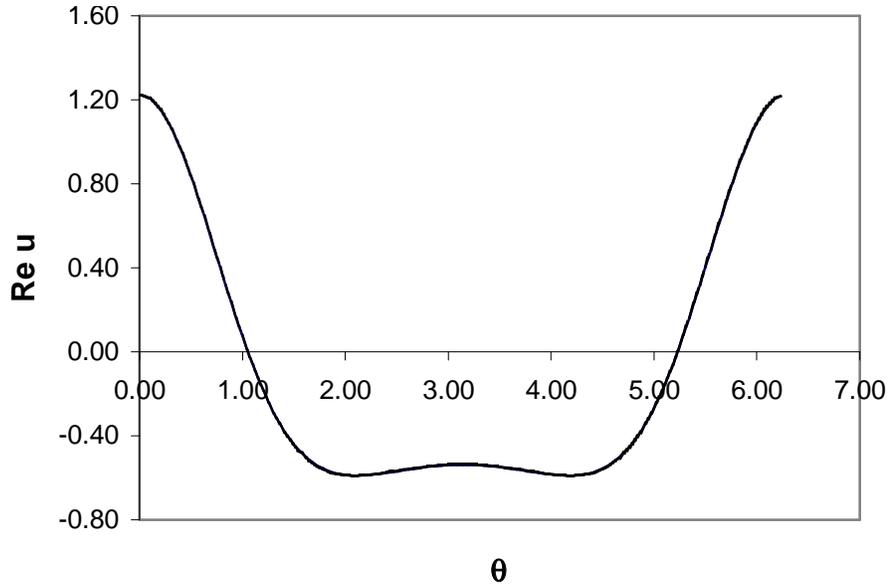}
\caption{Real part of the solutions for configurations $Q_1$ (solid line)
and $Q_2$ on the circle $S$
for $k_0=3$, $\Phi(Q_2)=.00012$.}
\end{figure}

\begin{figure}[tb]
\vspace{5pc}
\includegraphics*{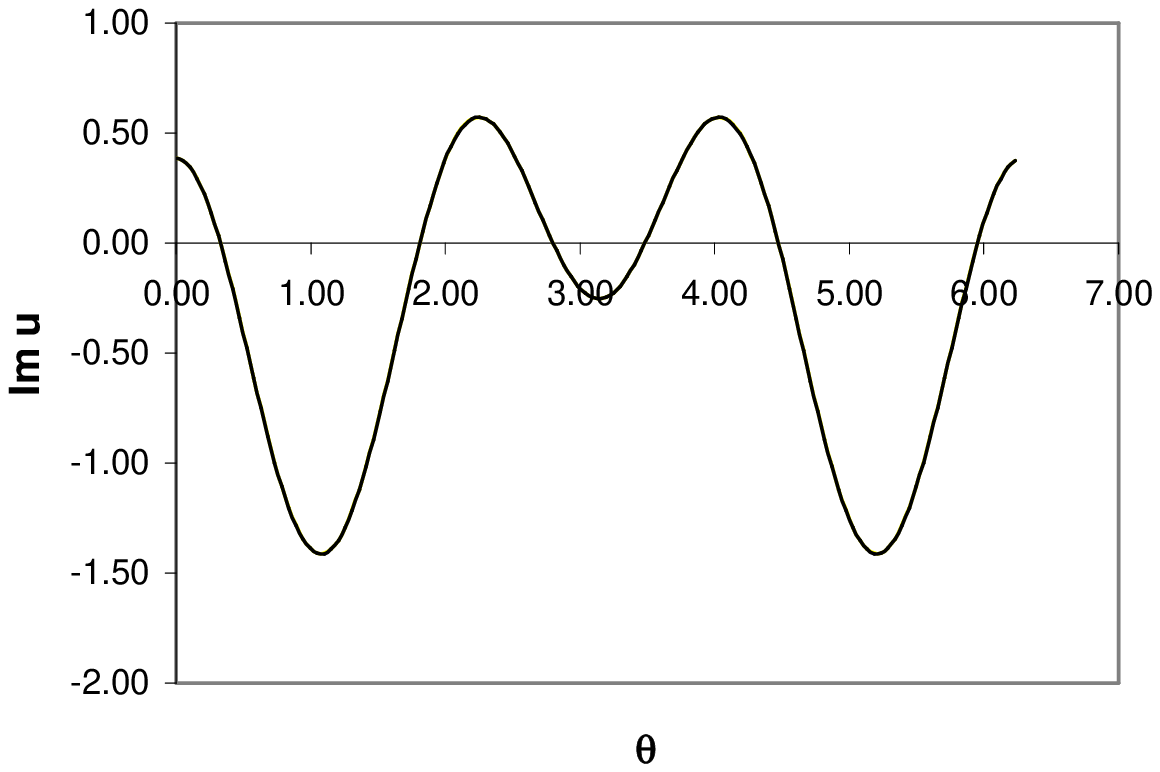}
\caption{Imaginary part of the solutions for configurations $Q_1$ (solid line)
and $Q_2$ on the circle $S$
for $k_0=3$, $\Phi(Q_2)=.00012$.}
\end{figure}

On the other hand, the situation is quite different if we allow the
scatterer to be probed with waves of multiple frequencies.

Figures 3 and 4 show the real and imaginary parts for the same
configurations $Q_1$ and $Q_2$ when the free space wave number $k_0$ is
equal to $10.0$. Then $\Phi(Q_2)=1.4307$. It is, of course, possible
 that there are configurations
undistinguishable at this frequency, but, combining the output for
several frequencies we can hope to achieve a reasonable recovery of the
original scatterer. We show in the subsequent sections, that it is, indeed,
the case. While there are many theoretical questions concerning the
best, or a reasonable choice of frequencies, uniqueness for the IPM,
stability estimates, etc., this work indicates the practicality of the
multifrequency approach.

\begin{figure}[tb]
\vspace{5pc}
\includegraphics*{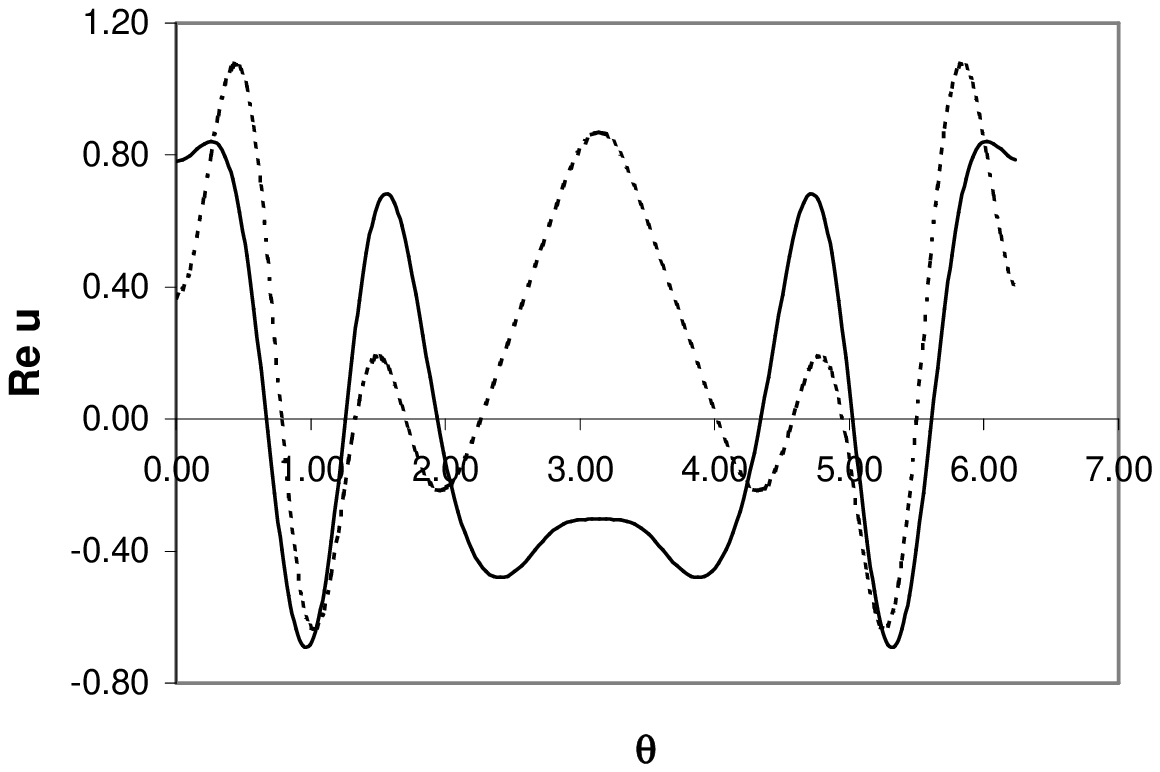}
\caption{Real part of the solutions for configurations $Q_1$ (solid line)
and $Q_2$ on the circle $S$
for $k_0=10$, $\Phi(Q_2)=1.4307$.}
\end{figure}

\begin{figure}[tb]
\vspace{5pc}
\includegraphics*{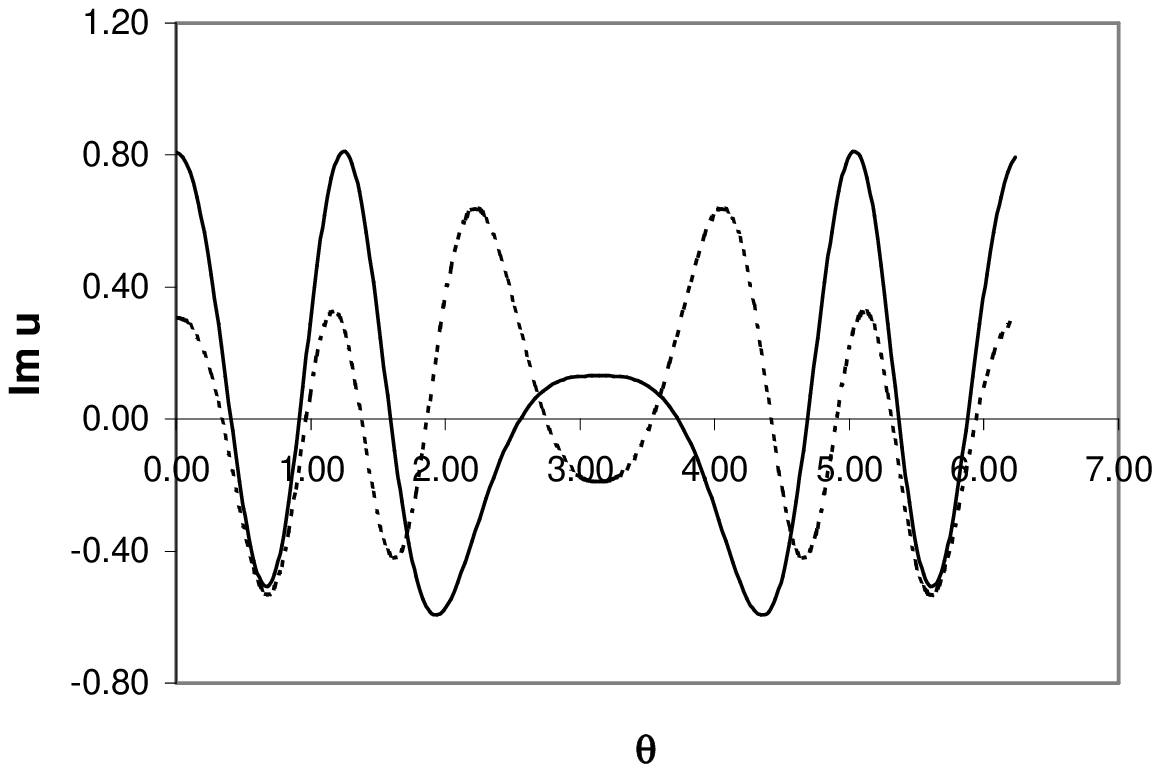}
\caption{Imaginary part of the solutions for configurations $Q_1$ 
 (solid line) and $Q_2$ on the circle $S$
for $k_0=10$, $\Phi(Q_2)=1.4307$.}
\end{figure}

To illustrate this point further, let  $P$ be the set of three free space wave
numbers $k_0^{(p)}$  chosen to be

\begin{equation}
P=\{3.0,\ 6.5,\ 10.0\}\,.
\end{equation}

Figure 5 shows the profile of the functional $\Phi$ as a function of the
variable $r\,,\, 0.1\leq r \leq 0.6$ in the configurations $q_r$ with

\[
n(x)=\begin{cases} 
0.49 & 0\leq |x| < r\\
9.0 & r \leq |x| < 0.6\\
1.0 & 0.6\leq |x| \leq 1.0
\end{cases}
\]

The best fit to data functional exhibits a sharp minimum at $r=.4$, thus
there is a hope to identify the sought configuration.

\begin{figure}[tb]
\vspace{5pc}
\includegraphics*{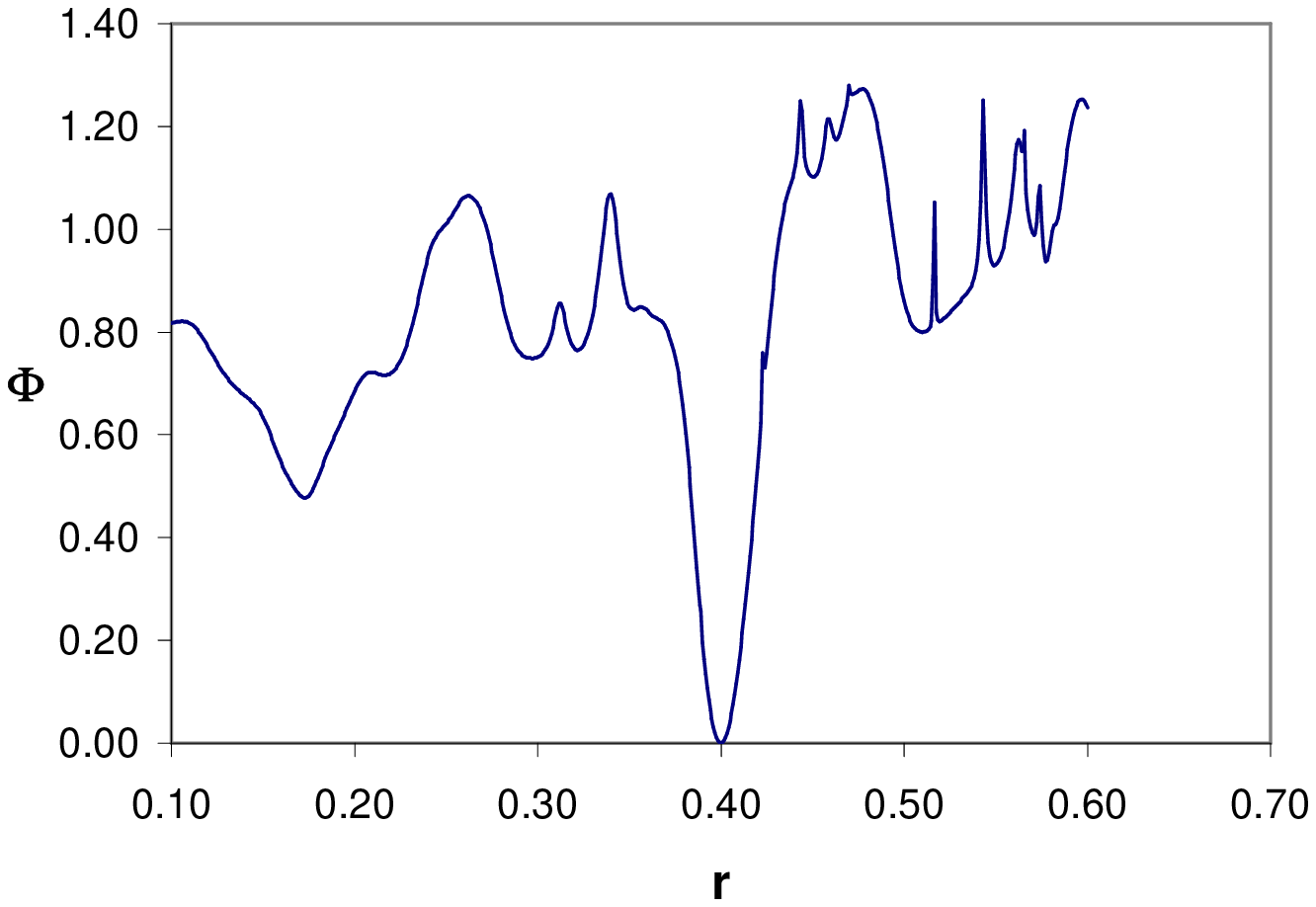}
\caption{Best fit profile for the configurations $q_r$; Multiple frequencies 
$P=\{3.0,\ 6.5,\ 10.0\}$.}
\end{figure}

\section{Local Minimization Methods}

Using the best fit to data functional $\Phi$ defined in (3.5), 
 the IPM is reduced to a restrained minimization  
over the admissible set $A_{adm}$,
defined in (3.6) and (3.7). It is well known that  a
multidimensional minimization is an extremely difficult problem,
unless the objective function is "well behaved". The most important
quality of such a cooperative function is the presence of just a few
local minima.
Unfortunately, this is, decidedly, not the case in many
applied problems, and, in particular, for the problem under the consideration.

Figure 5 shows that our objective function $\Phi$ has many
local minima even along this arbitrarily chosen one dimensional 
cross-section of the admissible set. There are sharp peaks and large
gradients.
Consequently, the gradient based methods (see \cite{bre},
\cite{des},\cite{fletcher},\cite{hestenes},\cite{jac},\cite{pol}) would not be successful
for a significant portion of this region. It is also appropriate to
notice that the dependency of $\Phi$ on its arguments is highly
nonlinear. Thus, the gradient computations have to be done numerically, which
makes them computationally expensive. More
importantly, the gradient based minimization methods (as expected)
perform poorly for these problems. 
These complications are avoided by considering conjugate gradient type algorithms
 which do not require the
knowledge of the derivatives at all. One such method is the Powell's method.

For an $N$-dimensional space the method can be described as follow (see
\cite{bre}).

\subsubsection*{Powell's Method}

\begin{enumerate}

\item  Initialize the set of directions $u_i$ to the basis vectors
\[
u_i=e_i\,,\quad i=1,2,\dots,N\,.
\]

\item  Save your starting position as $Q_0$.
\item  For $i=1,\dots,N$ move $Q_{i-1}$ to the minimum along the
direction $u_i$ and call this point $Q_i$.
\item  For $i=1,\dots,N$, set $u_i=u_{i-1}$.
\item  Set $u_N=P_N-P_0$.
\item  Move $P_N$ to the minimum along direction $u_N$ and call this
point $P_0$.

\end{enumerate}

It can be shown that an iteration of this procedure produces a set $u_i$
of mutually conjugate directions, provided, as usual, that the objective
function is quadratic. It also implies a quadratic convergence for nearly
quadratic functions. The main difficulty here is that the obtained set
of conjugate directions tends to become "folded up", that is linearly
dependent. However, as noted in \cite{bre}, the set of directions $u_i$ can
be reset to the basis vectors $e_i$ after every $N$ or $N+1$ iterations
of the basic procedure.

As explained in the next Section, we leave the global exploration of the
admissible set to global minimization methods. A local minimization is
used to explore an immediate vicinity of the initial configuration $Q\in
\rc^{2M}$. 
With this goal in mind, given a configuration $Q\in\rc^{2M}$
and a direction $u$ in $\rc^{2M}$ we seek a minimum of $\Phi$ along this
direction (by a bisection or a Golden Rule method) by restricting the probed
points (at every minimization step) to the admissible set
and by keeping them within a certain distance from the initial minimization point. This
distance $d_{min}$ is determined a priori to be a percentage of the
characteristic length of $A_{adm}$.

More precisely, the "turtle" one-dimensional minimization is done as
follows.  

\subsubsection*{One-dimensional minimization}
\begin{enumerate}

\item  Let the starting position be $Q_0$.
\item  Move from $Q_0$ along the given direction $u$ by the distance
$d_{min}$ to obtain $Q_1\in A_{adm}$. 
\item  Find the minimum of $\Phi$ on the interval $[Q_0,Q_1]$.

 If the minimum is attained inside the interval, then stop.

If the minimum is attained at $Q_0$, then reverse the direction.

If the minimum is attained at $Q_1$, then rename $Q_0=Q_1$, and repeat
the procedure.

\end{enumerate}

 We have used Brent's minimization Method
\cite{bre} for the one-dimensional minimization in the step 3.
This way the local minimum closest (the resolution is set up by $d_{min}$) to
  the starting
configuration $Q_0$ is determined. The choise of $d_{min}$ has to be
balanced between the desire to explore the fine structure of the
objective function and the computational costs.

 Now we can describe our Basic Local Minimization
Method in $\rc^{2M}$. The above "turtle" one-dimensional minimization
procedure is used in all the minimization steps below.

\subsubsection*{Basic Local Minimization Method}
\begin{enumerate}

\item  Initialize the set of directions $u_i$ to the basis vectors
\[
u_i=e_i\,,\quad i=1,2,\dots,2M\,.
\]

\item  Save your starting position as $Q_0$.
\item  For $i=1,\dots,2M$ move from $Q_0$ along the direction $u_i$ 
 to find the point of minimum $Q_i^t$.
\item  Reindex the directions $u_i$, so that (for the new indices) $\Phi(Q_1^t)\leq \Phi(Q_2^t)
\leq,\dots,\Phi(Q_{2M}^t)\leq\Phi(Q_0)$.

\item  For $i=1,\dots,2M$ move $Q_{i-1}$ to the minimum along the
direction $u_i$ and call this point $Q_i$.
\item  Set $v=Q_{2M}-Q_0$.
\item  Move $Q_{2M}$ to the minimum along direction $v$ and call this
point $Q_0$.
\item Repeat the above steps till a stopping criterion is satisfied.
\end{enumerate}

Note, that we use the temporary points of minima $Q_i^t$ only to rearrange the
initial directions $u_i$ in a different order. This method falls within
the category of the Powell's minimization methods, and, as mentioned
above, produces conjugate directions and a quadratic convergence for
nearly quadratic functions.

Still another refinement of the above algorithm has turned out to be necessary
to produce a successful minimization. Since the dimension $2M$ of the
minimization space was chosen a priori to be larger than $2N$, where $N$
is the (unknown) number of layers in the original scatterer, we expect
that the sought point of minimum will be located in a lower dimensional
subspace of the minimization space $\rc^{2M}$. This information
available from the specific structure of our minimization problem
appears to be nontrivial. Suffices to say, that all of our numerical
experiments described in Section 6 have failed without the following (space dimension) 
"reduction" procedure. The main idea behind it is to conduct the local minimization
searches in as low-dimensional subspaces as possible. It is specific to the inverse
scattering problem for multilayer scatterer.

If two adjacent layers
have  close refraction coefficients in the sense that the objective functional $\Phi$
is not changed much when the two layers assigned the same refraction coefficient, then these
two layers can be replaced with just one occupying their place. The minimization problem becomes
constrained to a lower dimensional subspace of $\rc^{2M}$ and the local
minimization is done in this subspace. A similar procedure was used by us
in \cite{gutmanramm} for the search of small subsurface objects.

\subsubsection*{Reduction Procedure}

Let $\epsilon_r$ be a positive number. 
\begin{enumerate}
\item  Save your starting configuration $Q_0=(r_1,r_2,\dots,r_M,n_1,n_2,\dots,n_M)$
 and $\Phi(Q_0)$. Let the $M+1$-st layer be $D_{M+1}=\{r_M\leq |x| \leq
R\}$ and $n_{M+1}=k_0^2$.

\item  For $i=2,\dots,M+1$ replace $n_{i-1}$ in the layer $D_{i-1}$ by $n_i$. Compute
$\Phi$ at the new configuration  $Q_i^d$, and the difference
$c_i^d=|\Phi(Q_0)-\Phi(Q_i^d)|$.

\item  For $i=1,\dots,M$ replace $n_{i+1}$ in the layer $D_{i+1}$ by $n_i$. Compute
$\Phi$ at the new configuration  $Q_i^u$, and the difference
$c_i^u=|\Phi(Q_0)-\Phi(Q_i^u)|$.

\item  Find the smallest among the numbers $c_i^d$ and $c_i^u$.
If this number is less than $\epsilon_r\Phi(Q_0)$, then adjust the
refraction coefficient to $n_i$ in the "down" or "up" layer accordingly.
Replace the two adjacent layers with one occupying their place, and
renumber the layers.

\item Repeat the above steps till no further reduction in the number of
layers is occurring.

\end{enumerate}

Note, that an application of the Reduction Procedure may or may not
result in the actual reduction of layers.

Finally, the entire Local Minimization Method {\bf (LMM)} consists of the
following:

\subsubsection*{Local Minimization Method (LMM)}
\begin{enumerate}

\item  Let your starting configuration be
$Q_0=(r_1,r_2,\dots,r_M,n_1,n_2,\dots,n_M)$.

\item  Apply the Reduction Procedure to $Q_0$, and obtain a reduced configuration
$Q_0^r$ containing $M^r$ layers.

\item  Apply the Basic Minimization Method in $A_{adm}\bigcap \rc^{2M^r}$ 
with the starting point $Q_0^r$, and obtain a configuration $Q_1$.

\item  Apply the Reduction Procedure to $Q_1$, and obtain a final reduced configuration
$Q_1^r$.

\end{enumerate}

\section{Global Minimization Methods}

Given an initial configuration $Q_0$ a local minimization method finds a
local minimum near $Q_0$. On the other hand, global minimization methods
explore the entire admissible set to find a global minimum of the
objective function. While the local minimization is, usually,
deterministic, the majority of the global methods are probabilistic in
their nature. There is a great interest and activity in the development of efficient
global minimization methods, see e.g. \cite{biegler},\cite{bomze}. Among them are the
simulated annealing method (\cite{kgv},\cite{kir}), various genetic algorithms \cite{hh},
 interval method,
TRUST method (\cite{bar},\cite{bpr}), etc. As we have already mentioned before, the best fit to
data functional $\Phi$ has many narrow local minima. In this situation
it is exceedingly unlikely to get the minima points by chance alone.
Thus our special interest is for the minimization methods, which combine
a global search with a local minimization. In \cite{gutmanramm} we 
developed such a method (the Hybrid Stochastic-Deterministic Method),
and applied it for the identification of small subsurface particles,
provided a set of surface measurements. The HSD method could be
classified as a variation of a genetic algorithm with a local search
with reduction. 
In this paper we consider the performance of two
algorithms: Deep's Method, and Rinnooy Kan and Timmer's 
Multilevel Single-Linkage Method. Both combine a global and a local search 
to determine a global minimum. Recently these methods have been applied
to a similar problem of
the identification of  particles from their light scattering characteristics in \cite{zub}.
Unlike \cite{zub}, our experience shows that
Deep's method has failed consistently for the type of problems we are
considering. See \cite{deep} and \cite{zub} for more details on Deep's
Method.

\subsubsection*{Multilevel Single-Linkage Method (MSLM)}

Rinnooy Kan and Timmer in \cite{rkt1} and \cite{rkt2} give a detailed
description of this algorithm. Zakovic et. al. in \cite{zub} describe in detail an
experience of its application to an inverse light scattering problem.
They also discuss different stopping criteria for the MSLM. 
Thus, we only give here a shortened and an informal description of this method and of its
algorithm. 

In a pure {\bf Random Search} method a batch $H$ of $L$ trial points is generated in
$A_{adm}$ using a uniformly distributed random variable. Then a local
search is started from each of these $L$ points. A local minimum with
the smallest value of $\Phi$ is declared to be the global one.

A refinement of the Random Search is the {\bf Reduced Sample Random Search} method.
Here we use
only a certain fixed fraction $\gamma<1$ of the original batch of $L$
points to proceed with the local searches. This reduced sample $H_{red}$ of
$\gamma L$ points is chosen to contain the points with the smallest
$\gamma L$ values of $\Phi$ among the original batch. The local searches
are started from the points in this reduced sample.

Since the local searches dominate the computational costs, we would like
to initiate them only when it is truly necessary. Given a critical
distance $d$ we define a cluster to be a group of points located within
the distance $d$ of each other. Intuitively, a local search started
from the points within a cluster should result in the same local
minimum, and, therefore, should be initiated only once in each cluster.

Having tried all the points in the reduced sample we have an information
on the number of local searches performed and the number of local minima
found. This information and the critical distance $d$ can be used to
determine a statistical level of confidence, that all the local minima
have been found. The algorithm is terminated (a stopping criterion is satisfied) 
if an a priori level of confidence is reached.

If, however, the stopping criterion is not satisfied, we perform another
iteration of the MSLM by generating another batch of $L$ trial points.
Then it is 
combined with the previously generated batches to obtain an enlarged batch $H^j$ of 
$jL$ points (at iteration $j$), which leads to a
reduced sample $H^j_{red}$ of $\gamma jL$ points. The
critical distance $d$ is reduced to $d_j$, (thus, the cluster's size is
redefined), a local minimization is attempted once within each cluster,
the information on the number of local minimizations performed and the
local minima found is used to determine if the algorithm should be
terminated, etc.

The following is an adaptation of the MSLM method to the inverse scattering problem
presented in Sections 2 and 3, with all the relevant notations. The LMM local
minimization method introduced in the previous Section is used here to
perform local searches.

\subsubsection*{MSLM}
(at iteration $j$).
\begin{enumerate}

\item  Generate another batch of $L$ trial points (configurations) from a random
uniform distribution in $A_{adm}$. 
Combine it with the previously generated batches to obtain an enlarged batch $H^j$ of 
$jL$ points. 

\item Reduce $H^j$ to the reduced sample $H^j_{red}$ of $\gamma jL$
points, by selecting the points with the smallest $\gamma jL$ values of $\Phi$
in $H^j$.

\item Calculate the critical distance $d_j$ by

\[
d_j^r=\pi^{-1/2}\left( \Gamma\left(1+\frac{M}{2}\right)R^M\frac{\sigma\ln jL}
{jL}\right)^{1/M}\,,
\]
\[
d_j^m=\pi^{-1/2}\left( \Gamma\left(1+\frac{M}{2}\right)(n_{high}-n_{low})^M\frac{\sigma\ln jL}
{jL}\right)^{1/M}\,.
\]
\[
d_j=\sqrt{(d_j^r)^2+(d_j^n)^2}
\]

\item Order the sample points in $H^j_{red}$ so that
$\Phi(Q_i)\leq\Phi(Q_{i+1})\,, i=1,2,\dots,\gamma jL$.
For each value of $i$, start the local minimization from $Q_i$, unless
there exists an index $k<i$, such that $\|Q_k-Q_i\|\leq d_j$. Ascertain if the
result is a known local minimum.

\item Let $K$ be the number of local minimizations performed, and $W$ be
the number of different local minima found. Let
\[
W_{tot}=\frac{W(K-1)}{K-W-2}
\]
The algorithm is terminated if

\begin{equation}
W_{tot}<W+0.5\,.
\end{equation}

\end{enumerate}

Here $\Gamma$ is the gamma function, and $\sigma$ is a fixed constant.

A related algorithm (the Mode Analysis) is based on a subdivision of the
admissible set into smaller volumes associated with local minima. This
algorithm is also discussed in \cite{rkt1} and \cite{rkt2}. From the
numerical studies presented there, the authors deduce their preference
for the MSLM.

\section{Numerical Results}
Introducing polar coordinates in (2.3) and separating the variables,
equations for the total field $u(x)$ become
\[
u_1(x)=\sum_{l=-\infty}^{\infty}a_{1,l}J_l(k_1|x|)e^{il\theta}
\]
for $x\in D_1$,

\[
u_m(x)=\sum_{l=-\infty}^{\infty}(a_{m,l}J_l(k_m|x|)+b_{m,l}Y_l(k_m|x|))e^{il\theta}
\]
for $x\in D_l,\,l=2,\dots,N$, and

\[
u(x)=e^{ik<x,\nu>}+\sum_{l=-\infty}^{\infty}A_lH_l^{(1)}(k_0|x|)e^{il\theta}
\]
for $x\in D : r_N\leq |x|\leq R$. Here $J_l, Y_l$  are the
Bessel functions of the first and second kind,  $H_l^{(1)}$ is the
Hankel function of the first kind, and $\nu$ is the direction vector of
the incident wave. Since
\[
e^{ik<x,\nu>}=\sum_{l=-\infty}^{\infty}i^lJ_l(k_0|x|)e^{il\theta}
\]
for $\nu=(1,0)$, the above equations and the conditions of continuity
form a system of equations from which the field $u(x)$ can be calculated
on the circle $S$. This solves the direct problem (2.2)-(2.3). Other methods of
solution for such problems are known as well, see e.g. \cite{hu},\cite{schuster}.
 Solving the direct problem
for the set $P$  of three free wave numbers $k_0^{(p)}$ (see (3.8)), we obtain
the total fields $u^{(p)}(x)$.
Their 
restrictions to $S$ give the (simulated) data $g^{(p)}(\theta)$.

Our approach to the inverse problem IPM (see Section 2) is to recast
it in the best fit to data form (3.5), and to minimize the objective functional $\Phi$
 over $A_{adm}$. 
We have tested  Deep's global minimization method, the Multilevel 
Single-Linkage Method, and a Reduced Sample Random Search method.
Each method was tried for three different original configurations $Q_0$
described below. The data $g^{(p)}(\theta)$ was computed at 120 angles
$\theta_l=2\pi l/120, l=1,2,\dots, 120$, and $\Phi$ was evaluated according to (3.4) and (3.5).
This data was used with three different
noise levels $\delta=0.00, 0.03$ and $0.10$. More precisely, for the
uniformly distributed on $[0,1]$ random variable $z$

\[
g_{\delta}(\theta)=g(\theta)+\delta\|g\|(2z-1)(1+i)
\]

for  the noise level $\delta$.

\bigskip
Since our goal was to test the applicable algorithms, the values for the
refraction coefficients, the size, the wave numbers, etc.,  were chosen arbitrarily at this time, 
that is without a regard 
for their possible physical relevancy. 
The original configurations are:


\paragraph{Configuration $Q_0^{(1)}$}
This is a one layer cylinder $Q_0^{(1)}=(0.72,4.2025)$ with $N=1$ and $R=1.0$, see
(3.2), that is the refraction coefficient is defined by
\[
n(x)=\begin{cases} 4.2025 & 0\leq |x| < 0.72\\
1.0 & 0.72\leq |x| \leq 1.0
\end{cases}
\]
\paragraph{Configuration $Q_0^{(2)}$}
This is a two layer cylinder $Q_0^{(2)}=(0.4,0.6,0.49,9.0)$ with $N=2$ and $R=1.0$,
that is
\[
n(x)=\begin{cases} 
0.49 & 0\leq |x| < 0.4\\
9.0 & 0.4\leq |x| < 0.6\\
1.0 & 0.6\leq |x| \leq 1.0
\end{cases}
\]
\paragraph{Configuration $Q_0^{(3)}$}
Three layer cylinder $Q_0^{(3)}=(0.3,0.7,0.8,4.0,25.0,9.0)$ with $N=3$ and $R=1.0$,
that is
\[
n(x)=\begin{cases} 
4.0 & 0\leq |x| < 0.3\\
25.0 & 0.3\leq |x| < 0.7\\
9.0 & 0.7\leq |x| < 0.8\\
1.0 & 0.8\leq |x| \leq 1.0
\end{cases}
\]

To identify these configurations we applied the global minimization
methods of Section 5. In each one we let $M=4\,,R=1.0$. 
A priori bounds for the refraction coefficients were chosen to be
$n_{low}=.04$ and $n_{high}=30.25$.
Minor
modifications to the description of the methods in Section 5 were
introduced for the purpose of computational simplification. In
particular, the minimization was done in $\sqrt{n_m}$ rather than in $n_m$
as stated there. This results in a rescaling of the admissible set. In
each case, after a global minimum $Q_{min}$ was determined, the error of
the identification 

\begin{equation}
\epsilon_{err}= \frac{\int_D|n_{min}(x)-n(x)|}{\int_D n(x)}\,,
\end{equation}

where $n(x)$ is the refraction coefficient of the original configuration
$Q_0$,
was computed to determine if the
identification was successful. We distinguished between the two levels of a successful
identification: $\epsilon_{err}<0.01$ and $\epsilon_{err}<0.1$.

\subsubsection*{Identification by Deep's Method(\cite{deep},
\cite{zub})}
Each test of the method consisted in 100 independent runs. Since $M=4$
the minimization was done in $\rc^8$. As we have already mentioned, the
method failed every time. It seems, that the local minimization phase of Deep's
method (minimization over randomly selected parabolas) is not extensive
enough to identify narrow local minima present in this problem. Also,
the method does not use the Reduction Procedure (see Section 4), which,
we think, is another reason for its failure. As in \cite{zub} we have also
observed the cycling of the algorithm.

\subsubsection*{Identification by Reduced Sample Random Search Method} This method
is presented in Section 5 in the subsection on the Multilevel Single-
Linkage Method. The Local Minimization Method with the Reduction
Procedure (as described in Section 4) was used in the local minimization
phase. Chosen parameters $L=15000$ and $\gamma=0.01$ the performance of
this method is the same as the Multilevel Single-Linkage Method. In
fact, $L=15000$ is exactly the sample size in MSLM at it termination in our experiments.
Since MSLM has the great advantage of a self-contained statistical
stopping criteria (and from which the number 15000 was determined in the
first place), it is, clearly, a preferred method.

\subsubsection*{Identification by Multilevel Single-Linkage Method} 

We have attempted to identify all 3  original
configurations $Q_0^{(1)}, Q_0^{(2)}$ and $Q_0^{(3)}$. Each one with no
noise in the data ($\delta=0.00$) as well as with noise levels
$\delta=0.03$ and $\delta=0.10$. Each of the 9 tests consisted of 10
independent runs. It took about 60 to 80 minutes on average to complete one run
on a 333 MHZ PC. We used  $M=4\,,R=1.0\,,\gamma=0.01$ and the sample size
$L=200$. The parameter $\sigma$ was chosen to be equal to $1.0$. Value
$\sigma=4.0$ was used in \cite{rkt2}, and $\sigma=1.9$ in \cite{dixonjha}.
As in Deep's Method above, 
a priori bounds for the refraction coefficients were chosen to be
$n_{low}=.04$ and $n_{high}=30.25$. The value $\epsilon_r=0.1$ was used in
the Reduction Procedure (see Section 4) during the local minimization phase.

As in other works on the clustering algorithm, we have found the
stopping rule (5.1) to be unsatisfactory. In our experience the difference
$W_{tot}-W$, while slightly decreasing with the number of performed
minimizations, has   quickly stabilized around the value of 5.
Thus, the stopping criterion (5.1)

\[
W_{tot}<W+0.5\,.
\]

could not be attained. This issue has been discussed in \cite{dixonjha} and 
\cite{boender},
where a different stopping rule was suggested for functions with large
number of local minima. Since the Bayesian stopping rule reflects the
level of
confidence in finding all the local minima, a relaxation of (5.1) would
mean a smaller level of confidence, which still may be acceptable to
assure that the global minimum is found among already performed local
minimizations. We have chosen to replace (5.1) with

\begin{equation}
W_{tot}<W+0.5\quad \text{or} \quad W_{tot}<(1+\epsilon_{tot})W\,,
\end{equation}
where $\epsilon_{tot}=0.03$.

As before

\[
W_{tot}=\frac{W(K-1)}{K-W-2}\,,
\]
where $K$ is the number of local minimizations performed, and $W$ is
the number of different local minima found. 
Thus, the MSLM algorithm is terminated if (6.2) is satisfied. In our
numerical experiments we have got the following average values
$K=5000\,,W=150$, and $W_{tot}=155$.


\begin{table}
\caption{Identification by MSLM. Original configuration $Q_0^{(1)}$.}
\begin{tabular*}{\textwidth}{@{\extracolsep{\fill}}rrrrr}
\hline
 & & \multicolumn{2}{c}{Success rate$^1$} & 
\\
\cline{3-4}
Noise & Runs & $0.01$ & $0.1$ & Smallest $\Phi$  \cr
\hline
 $\delta$=0.00 & 10 & 8 & 10 & 0.0000 \cr
 $\delta$=0.03 & 10 & 9 & 10 & 0.0016 \cr
 $\delta$=0.10 & 10 & 10 & 10 & 0.0178 \cr
\hline

\end{tabular*}

\end{table} 

\footnotetext[1]{ Identification is successful if
$\epsilon_{err}<0.01$, 
  or $\epsilon_{err}<0.1$ correspondingly, see (6.1).}

An example of a successful ($\epsilon_{err}<0.1$) identification
for $Q_0^{(2)}$ and $\delta=0.10$ is shown on Figure 6. The identified
configuration is a two layer cylinder

\[
 Q_{id}=(0.3966,0.5943,0.4684,9.203)\,,
\]

with $\Phi(Q_{id})=0.0367,\epsilon_{err}=0.0480$. That is

\[
n(x)=\begin{cases} 
0.4684 & 0\leq |x| < 0.3966\\
9.203 & 0.3966 \leq |x| < 0.5943\\
1.0 & 0.5943 \leq |x| \leq 1.0
\end{cases}
\]

\begin{figure}[tb]
\vspace{5pc}
\includegraphics*{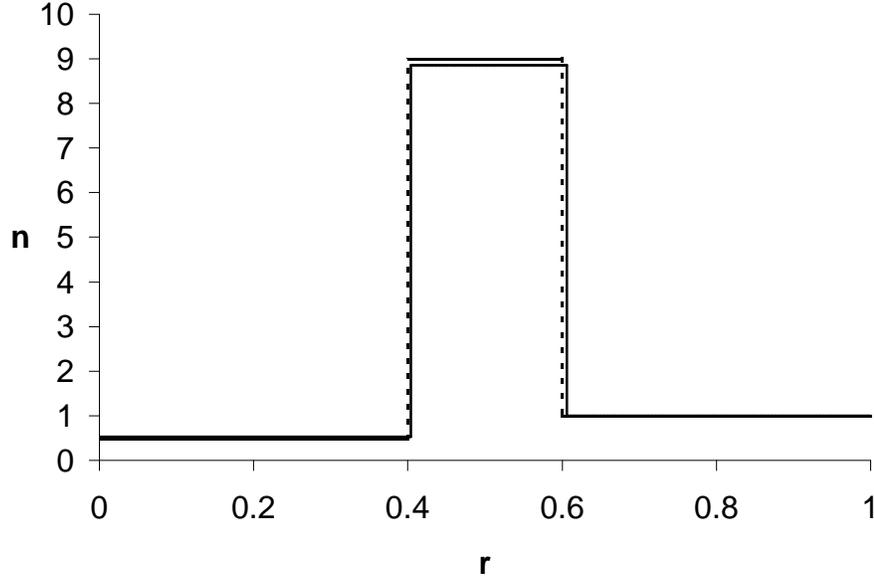}
\caption{Refraction coefficients $n(x)$ for the original $Q_0^{(2)}$
 and the identified $Q_{id}$ (solid line) configurations.
 Data noise level $\delta=0.10$, 
$\Phi(Q_{id})=0.0367,\epsilon_{err}=0.0480$.}
\end{figure}


\begin{table}
\caption{Identification by MSLM. Original configuration $Q_0^{(2)}$.}
\begin{tabular*}{\textwidth}{@{\extracolsep{\fill}}rrrrr}
\hline
 & & \multicolumn{2}{c}{Success rate$^1$} & 

\\
\cline{3-4}

 Noise & Runs & $0.01$ & $0.1$ & 
Smallest $\Phi$  \cr
\hline
 $\delta$=0.00 & 10 & 10 & 10 & 0.0000 \cr
 $\delta$=0.03 & 10 & 1 & 10 & 0.0034 \cr
 $\delta$=0.10 & 10 & 2 & 9 & 0.0362 \cr 
\hline
\end{tabular*}

\end{table}


\begin{table}
\caption{Identification by MSLM. Original configuration $Q_0^{(3)}$.}
\begin{tabular*}{\textwidth}{@{\extracolsep{\fill}}rrrrr}
\hline
 & & \multicolumn{2}{c}{Success rate$^1$} & 
\\
\cline{3-4}

 Noise & Runs & $0.01$ & $0.2$ & 
Smallest $\Phi$  \\
\hline
 $\delta$=0.00 & 10 & 2 & 7 & 0.0000 \\
 $\delta$=0.03 & 10 & 0 & 5 & 0.0071 \\ 
 $\delta$=0.10 & 10 & 0 & 5 & 0.0541 \\ 
\hline
\end{tabular*}

\end{table}

An example of a successful ($\epsilon_{err}<0.1$) identification
for $Q_0^{(3)}$ and $\delta=0.03$ is a three layer cylinder

\[
 Q_{id}=(0.3030,0.7067,0.8079,4.071,24.528,8.857)\,,
\]

with $\Phi(Q_{id})=0.00708,\epsilon_{err}=0.04125$. That is

\[
n(x)=\begin{cases} 
4.071 & 0\leq |x| < 0.3030\\
24.528 & 0.3030\leq |x| < 0.7067\\
8.857 & 0.7067\leq |x| < 0.8079\\
1.0 & 0.8079\leq |x| \leq 1.0
\end{cases}
\]

\section{Conclusions}
The inverse scattering problem IPM is the identification of a multilayered scatterer
by a set of observations on its boundary. Such problems have 
applications in science and engineering. In the case of
the weak scattering approximation, many such problems can be solved by a
linearized inversion. However, if the scattering is not weak, other
methods of solution need to be developed.  We
have illustrated in Section 3, that an inversion based on just one frequency
of the incident waves cannot be successful, since there are distinct
configurations, producing practically the same observations. Introducing
multiple frequencies, however, makes the inverse problem more amenable
to a solution.

In this paper the inverse
problem is transformed into the best fit to data minimization problem.
This
minimization is difficult, since the objective function is rugged and
has many narrow local minima. A promising way to treat such a
minimization is by a combination of global (probabilistic) and local
(deterministic) minimization methods. In this paper we examined various
local and global methods. Concerning the local minimization methods it
was shown, that the Local Minimization Method (LMM) of Section 4 was
successful, even where other considered methods failed. This method is a
variation of a conjugate directions method with no use of partial derivatives.
It has a quadratic
convergence near quadratically shaped minima. However, even this
method needs to be enhanced by a Reduction Procedure (Section 4).
This procedure helps the minimization to take an advantage
of the a priori available information, that the sought minima are likely
to be found in certain lower dimensional subspaces of the entire
minimization space.

For the global minimization part we considered Deep's method and the
Multilevel Single-Linkage Method. While Deep's method failed, the MSLM
was successful in many instances. It also has an important advantage of
having termination criteria establishing a level of confidence, that the
found minima contain the sought global minimum. Among the deficiencies
of the MSLM are its slow execution, and inconsistency and failture to identify some
configurations. There is still a problem in choosing an appropriate
stopping rule. Thus, the MSLM provides a benchmark, against which the performance other
methods can be judged and measured.

\begin{acknowledgment}
I would like to thank the referees for their valuable suggestions.
\end{acknowledgment}

\end{article}

\end{document}